\documentstyle[aps,pre,multicol,psfig,epsf,amssymb,graphicx,exscale,
fontenc,latexsym]{revtex}
\draft

\newcommand{\oo}{\infty}                 	  

\newcommand{\nsaw}{\nu_{\mbox{\tiny{SAW}}}}       
\newcommand{\nth}{\nu_{\mbox{\tiny{$\Theta$}}}}   
\newcommand{\fth}{\phi_{\mbox{\tiny{$\Theta$}}}}  
\newcommand{\Tth}{T_{\mbox{\tiny{$\Theta$}}}}     
\newcommand{\sigq}{\sigma_{\!q}}		  
\newcommand{\Prad}{P_{\!\mathit{rad}}}		  
\newcommand{\Nmin}{N_{\mbox{\tiny{\it MIN}}}}	  
\newcommand{\Nmax}{N_{\mbox{\tiny{\it MAX}}}}	  

\begin{document}
\title{Peculiar scaling of self-avoiding walk contacts}
\author{Marco Baiesi$^{1,*}$, Enzo Orlandini$^{1,*}$ 
and Attilio L.~Stella$^{1,2,*}$}
\address{$^1$ INFM-Dipartimento di Fisica,\\
Universit\`a di Padova, I-35131 Padova, Italy\\
$^2$ Sezione INFN, Universit\`a di Padova, I-35131 Padova, Italy }
\date{Received 12 March 2001} 
\maketitle

\begin{abstract}
The nearest neighbor contacts between the two halves of an $N$-site 
lattice self-avoiding walk offer an unusual example of scaling random
geometry: for $N \to \oo$ they are strictly finite in number 
but their radius of gyration $R_c$ is power law distributed 
$\propto R_c^{-\tau}$, 
where $\tau>1$ is a novel exponent characterizing universal behavior.
A continuum of diverging lengths scales is associated to the 
$R_c$ distribution.
A possibly super-universal $\tau=2$ is also expected for the 
contacts of a self-avoiding or random walk with a confining wall.
\bgroup
\egroup
\end{abstract}

 \begin{multicols}{2} \narrowtext

The self-avoiding walk (SAW) is a classical problem in statistical
mechanics, playing a central role in our understanding of
polymer statistics and intimately related to magnetic critical
phenomena and percolation~\cite{Vande}.
 In its most simple version the SAW amounts
to the statistical characterization of equally probable
single chain conformations, with no overlaps or intersections.
These conformations are made by $N-1$ successive 
nearest neighbor (nn) steps ($N$ sites) on a lattice in $d$ dimensions.
For $N \to \oo$, quantities like the radius of gyration with respect
to the center of mass of SAW configurations, $R_g$, have an average 
$\langle R_g \rangle \sim N^{\nsaw}$, where $\nsaw$
 is the SAW metric exponent.
When $d>1$ self-avoidance
does not prevent the conformations from involving close
approaches of different parts of the chain: 
here we generally count as contacts
the pairs of nn lattice sites which are visited,
not consecutively, by the SAW. The totality of such contacts
grows on average proportional to $N$~\cite{Kest} and possesses the same 
fractal dimension ($\equiv 1/\nsaw$) as the whole SAW.

In the present Letter we discuss so far unexplored
features of a particular subset of SAW contacts. 
Such features represent an unusual example of how scale invariance
can manifest itself in a finite, non-extensive portion of an infinite
fractal set. 
The scaling exponent $\tau$ of the probability distribution of
the gyration radius of such subset represents a novel characterization of
SAW universal behavior. 
The fact that $\tau>1$ implies the existence of a whole continuum of
diverging characteristic lengths in the SAW problem, in addition to the
length $\langle R_g \rangle$.

\begin{figure}[hbt]
\centerline{
\psfig{file=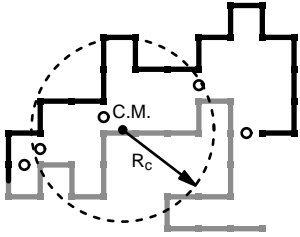,height=3.1cm}}
\vskip 0.3truecm
\caption{SAW in two dimensions: the contacts between its two 
halves (light and dark, respectively) are indicated by open circles. 
$\mathsf{C.M.}$ is the center of mass of these contacts and  
$\mathsf{R_c}$ their radius of gyration.}
\label{FIG1}
\end{figure}

The subset on which we focus here is that of the contacts between the
two halves of a SAW (Fig.~\ref{FIG1}). 
Counting such contacts alone allows to get rid of less interesting 
effects, which are extensive in $N$.
Being related to problems like network formation, transport or intramolecular
reactions, these contacts can be of particular interest for applications in 
which the two half-chains are made of different monomers, like for diblock
copolymers~\cite{copo}. 
When the chain represents a homopolymer, 
the subset we consider is particularly significant in 
relation to the effect of nearest neighbor interactions on the SAW~\cite{DI95}.
This is the case of models of 
the polymer $\Theta$ collapse~\cite{Vande,TJOW96,G97}, where an attractive
nn energy $\epsilon<0$ is associated to each contact.
In this case a Boltzmann factor $e^{-\epsilon/T}$ weighs each contact
occurring in a configuration at temperature $T$.
If for such model one counts only contacts between the two halves of the SAW, 
the average number of them is of the order
$N^0$ in the high $T$ regime, increases as
$N^{\phi}$, with $0< \phi <1$, at the $\Theta$ point, and scales as $N$ at low
$T$. 
$\phi$ turns out to coincide with the crossover exponent $\fth$ at $T=\Tth$, 
where $\langle R_g \rangle \sim N^{\nth}$, with $\nth \neq \nsaw$~\cite{DC1}. 
Similar behaviors occur if the attractive
interactions are associated exclusively with this subset of contacts,
and the model describes a diblock copolymer zipping
transition~\cite{DC1,DC2}. 
In different $T$ regimes of the $\Theta$ and zipping transitions the
contacts between the two halves 
of the SAW behave in a way analogous to the monomers adhering to a wall
in polymer adsorption~\cite{DC1}.
Most recently it has also been found that in $d=2$
the fractal dimensions of the contacts between the two half-chains
are the same at the homopolymer $\Theta$ point and at the diblock copolymer 
zipping transition~\cite{DC2}. Thus, intriguing universality aspects
can be expected to underlie the geometry of SAW contacts.

\begin{figure}[hbt]
\centerline{
\psfig{file=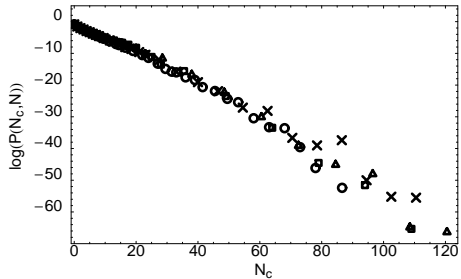,width=6.0cm}}
\vskip 0.2truecm
\caption{Histograms of $\ln[P(N_c,N)]$ for SAW on a cubic lattice, 
with $ N $ = 200 ({\footnotesize \mbox{$\bigcirc$}}), 
400 ({\large{$\times$}}), 800 ($\triangle$) and 1600 ($\square$).}
\label{FIG2}
\end{figure}

The contacts between the two halves of a polymer have already been
studied by renormalization group methods~\cite{MSc98}
in the high $T$ regime controlled by excluded volume.
The focus there was on the scaling of their  average 
number, $\langle N_c \rangle$, 
and precisely on the scaling correction exponent
describing its approach to a finite limit for $N\to \oo$. 
This finite limit shows that only a vanishing fraction of
the total number of contacts pertains to approaches of remote
portions of the chain.
In fact $\langle N_c \rangle$ gives only limited information on the
contact statistics. The full
probability distribution function (PDF) for $N_c$,
$P(N_c,N)$, could in principle have higher moments
diverging with $N\to \oo$, and we devote a first effort
to check this possibility, which is normally not
considered in polymer statistics. 
We perform several Monte Carlo simulations to investigate this PDF
and its moments in various dimensions and for $80 \leq N \leq 2000$.
Since the configurations with many contacts are very rare, we employ a
pruned-enriched Rosembluth method (PERM~\cite{G97}) tuned to increase the 
sampling of configurations with high number of contacts. 
This enables us to obtain a good statistics also in the tail
of the PDF, which for all $d$ turns out to be a negative
exponential without substantial dependence on $N$,
 as displayed in Fig.~\ref{FIG2} for SAW on cubic lattice.

\begin{figure}[hbt]
\centerline{
\psfig{file=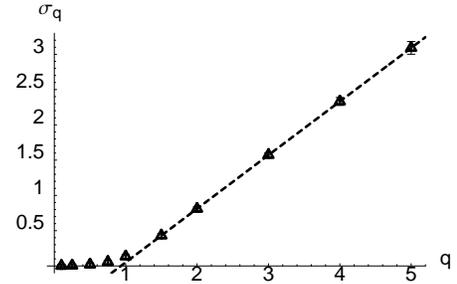,width=5.8cm}}
\vskip 0.2truecm
\caption{$\sigq$ vs $q$ and extrapolation of $\tau-1$ for $d=2$. 
We expect $\sigq=0$ for $q\leq \tau-1$.
Numerically a logarithmic divergence can not be 
easily distinguished from a power law one with $\sigq \gtrsim 0$.}
\label{FIG3}
\end{figure}

\begin{figure}[hbt]
\centerline{
\psfig{file=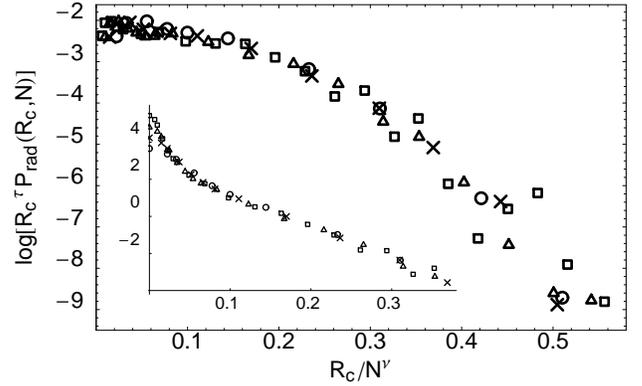,width=8.2cm}}
\vskip 0.2truecm
\caption{Collapses of $\ln[f(x)]$ (see the text) and $\ln[g(x)]$ (inset)
for SAW on cubic lattice, 
with $ N $ = 200 ({\footnotesize \mbox{$\bigcirc$}}), 
400 ({\large{$\times$}}), 800 ($\triangle$) and 1600 ($\square$).}
\label{FIG4}
\end{figure}

These results show that
this set of contacts is strictly finite in the $N\to \oo$ limit.
In spite of this, one can still ask whether these contacts have
interesting geometrical scaling properties,
not just reducing to those of a spatially bounded random set.
If we indicate by 
$\Prad(R_c,N)$ the cumulative PDF of $R_c$ (Fig.~\ref{FIG1})
over all SAW configurations, strict boundedness would mean that the moments 
$\langle R_c^q \rangle\sim \int dR_c\,R_c^q \Prad(R_c,N)$
remain finite, for $N \to \oo$, $\forall q$.
We extrapolate the moments in the form $\langle R_c^q \rangle
\sim N^{\sigq}$. The data are generated by sampling SAW of fixed length
with a Monte Carlo algorithm based on pivot moves~\cite{pivot}, which have been
 proved to be very efficient for SAW's~\cite{nu3d}.
Since the contacts between the two halves of the SAW are mainly 
located close to the
 junction point, local moves~\cite{local} are also attempted
 in this region. We consider hypercubic lattices and the FCC lattice in $d=3$. 
Contrary to what happens for $P(N_c,N)$, for all $d$ we find that 
$\sigq$ is positive and grows linearly with $q$, at sufficiently high $q$. 
The data are consistent with a behavior
\begin{equation}
\sigq= \left\{ \begin{array}{lcrr}
0 & \,\mathrm{if} &\, q \leq \tau-1 & \\
\nu\,[q-(\tau-1)] & \,\mathrm{if} & \, q>\tau-1 & ,
\end{array} \right.
\label{E1}
\end{equation}
where $\sigq=0$ may represent logarithmic divergences for $0<q\leq \tau-1$.
In $d=2$, for example (see Fig.~\ref{FIG3}), we find $\tau=1.93(2)$
and $\nu = 0.76(1)$, consistent with $\nu = \nsaw = 3/4$~\cite{Nie}.
The form (\ref{E1}) is compatible with a PDF having a scaling form
$\Prad(R_c,N)\simeq R_c^{-\tau} f(R_c/N^{\nu})$. 
That $\nu = \nsaw$ is quite plausible since 
$N^{\nsaw}$ is the only expected characteristic length in the problem.
However, as discussed below, $\Prad$ itself introduces a multiplicity of
new lengths scales for the SAW.  
The fact that the moments do not approach zero 
for $q < \tau-1$ (i.e.~$\sigq$ does not become negative) 
should be due to the circumstance that
$f(x)$ is not converging to zero for $x \to 0$. 
In other terms, the scaling function
$g$, if we write $\Prad\simeq N^{-\tau \nu} g(R_c/N^{\nu})$,
is singular for its argument approaching zero ($g(x)\sim x^{-\tau}$).
We verify these assumptions on the structure of $\Prad$
by scaling collapse plots for various $d$.
For example, Fig.~\ref{FIG4} shows the data collapse of $\ln[f(x)]$ for $d=3$.
A similar collapse plot for $\ln[g(x)]$ is reported in the inset.

It is indeed the singular character of $g(x)$ for $x \to 0$, 
which allows the exponent $\tau$ to take a nontrivial value $>1$, while
maintaining the zero-th, normalization moment of the PDF equal to $1$. 
The lower length cutoff $l$ (lattice spacing in this case)
is crucial to obtain a finite integral, in $dx=d(R_c/N^\nu)$,
of the PDF in the continuum limit, because the main contribution 
comes from small values of $x$, close to $x_-\equiv l/N^\nu$.  
This contribution has an $N$-dependence which compensates the diverging
factor $N^{- \nu (\tau-1)}$ extracted in front of the integral.

$\tau>1$ means that we can not associate to the $R_c$ PDF a unique
characteristic length. 
Indeed, putting $\xi_q \equiv {\langle R_c^q \rangle}^{1/q}$ we find 
$\xi_q \sim N^{\nu_q}$ with 
$\nu_q \equiv \sigma_q/q,$ for $q\in [\tau-1,+\oo)$.
This means that the self-similarity of contacts has an intrinsic 
multiscaling character.

The $\tau$ found here is a novel exponent for the SAW.
It is a measure of the spread of the region within
which one half of the chain 
feels the presence of  the other one in the SAW configurations.
A higher $\tau$ (see Table \ref{T1}) indicates more localized contacts.
Like the global geometry of the SAW, the spread of contacts is determined 
by the interplay between entropic and excluded volume effects.
 It is remarkable the non-monotonic
behavior of $\tau$, which takes minimum values
in $d=3$ and $d=4$, indicating these dimensionalities as the
optimal ones for a broad interpenetration of the two half-chains.
At high $d$ there is soon much room for the two branches of the
SAW to develop without giving rise to contacts,
and this tends to localize them more.
$\tau$ is rather high in $d=2$: we interpret this as a consequence
of the peculiar topology of the two-dimensional lattice,
which makes it more difficult than, e.g., in $d=3$ for the two half-chains
to approach each other at large length scales.
The dependence of $\tau$ on $d$, also above the upper critical
dimension $d_u = 4$, is a further indication of the peculiar novel
character of this exponent.

\begin{figure}[hbt]
\centerline{
\psfig{file=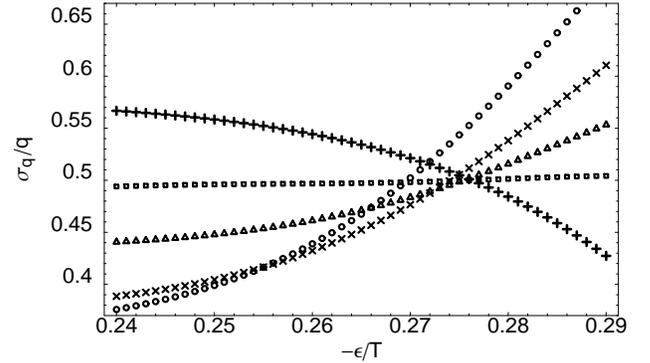,width=8.2cm}}
\vskip 0.2truecm
\caption{Extrapolation of $\sigq / q$ close to $\Tth$ in $d=3$, 
for $q=0.5$ ({\mbox{\footnotesize{$\bigcirc$}}}), $q=1$ ({\large{$\times$}}), 
$q=2$ $(\triangle)$ and $q=4$ $(\square)$. The crossings are
consistent with the expectation to find $\tau=1$ and 
$\sigq/q = \nth = 1/2$ right at $T=\Tth$.
The curve with the opposite trend ($+$) refers to the effective
 $\nu$ exponent of the SAW radius of gyration.}
\label{FIG5}
\end{figure}

In $d$=2 and $d$=3 we also investigate the behavior of the contacts between
the two half-chains in the presence of nn attractive interactions
($\Theta$ point model).
While for $T>\Tth$ the behaviors of their PDF's
appear the same as at  $T=\oo$, at the $\Theta$ point 
the moments of $P$ diverge as those of $\Prad$, while
$\tau$ becomes equal to $1$ and $\sigq / q = \nth$ for the latter PDF.
This suggests that the disappearance of the singular scaling
function in $\Prad$ could be a good criterion for locating the transition
point. This is illustrated in Fig.~\ref{FIG5}, referring to
the $\Theta$ point in $d=3$, simulated as in ref.~\cite{G97}
with chains up to $N=2000$. 
We collect data from two runs at $-\epsilon / T={0.25, 0.27}$ and we use the
multiple histogram method~\cite{hm} to calculate the moments in a surrounding
interval of temperatures. 
The result from the values of $q$ examined is $-\epsilon / \Tth = 0.274(4)$,
 consistent with accurate estimates by other methods: 
for example, $0.275(8)$ in~\cite{TJOW96} and $0.2690(3)$ in~\cite{G97}. 
Similar results are valid for models of the
diblock copolymer zipping transition~\cite{BOSf}.

For SAW with attractive interactions
the average number of contacts between the two half chains 
behaves in the various temperature ranges as the mean
number of SAW-wall contacts in a polymer adsorption transition.
Thus, it makes sense to check whether also in the adsorption process
the high $T$ ordinary regime is characterized by peculiar scaling
features of the kind described above. 
To this purpose we study the 
PDF's of the number and the radius of SAW-wall contacts
at $T= \oo$, where the wall exerts only a geometrical confinement
effect on the chain.
Here we consider as radius the mean 
distance $R_c^*$ of a contact from the point where the
SAW is grafted to the wall~\cite{n1}. Also in this case all the
moments of the PDF of $N_c$ appear to remain finite for
$N\to \oo$. 
The PDF's of $R_c$ show singular
scaling functions with an exponent $\tau \simeq 2$, 
in $d=2$ and $d=3$ (Table \ref{T2}).

We study also the case of RW confined by a wall, for which
we are able to compute exactly $\tau$.
Consider first a RW on a square lattice tilted of $45^\circ$ with respect
to the coordinate axes, in such a way that the nn 
of a site have coordinates $\{(1,1),\,(-1,1),\,(-1,-1),\,(1,-1)\}$. 
In this way a nn step moves the RW with nonzero and 
independent displacements along both
coordinate directions. This simplifies the calculations,
but we expect the final results to be valid for every lattice 
 model, because independence is asymptotically recovered for long RW.  
The generalization to $d$ 
dimensions gives $2^d$ nn vectors of the form
$(1,1,\ldots,1), (-1,1,\ldots,1),\ldots,(-1,-1,\ldots,-1)$ and, 
for example, in $d=3$ one obtains the BCC lattice. 
Let the walk start from the origin, near a $d-1$ dimensional hard wall
perpendicular to the $x$ coordinate:  the problem
in the $x$ direction is equivalent to a one dimensional 
RW which steps to nn sites with equal probability.
The probability to be again at $x=0$ after $n$ steps is given by
$P_n(0)=2^{-n} {n \choose n/2}$ .
The effect of the impenetrable wall is represented by forbidden sites
located at $\bar{x}=-1$.
So one has to subtract the probability to travel through this point.
The method of images~\cite{BaNi} in this simple case says that this is equal 
to the probability to go from $x=0$ to its mirror image with respect to 
$\bar{x}$, $x=-2$. So
$\bar{P}_n(0)=P_n(0)-P_n(-2)=P_n(0) 2(n+2)^{-1}$
is the probability to be on the surface. The $q$-th moment of
the average distance of a contact from the origin is thus
\begin{equation}
\langle R_c^*(N)^q \rangle \sim
	\frac{ \sum_{n=2,4,6,..}^N  \bar{P}_n(0) \big(n^{1/2}\big)^q}
	     { \sum_{n=2,4,6,..}^N  \bar{P}_n(0)}
\label{Rcq}
\end{equation}
where $n^{1/2}$ is the root mean square displacement after $n$ steps.
Using $n!\approx \sqrt{2 \pi}\, n^{n+1/2} e^{-n}$ for the binomials,
one recovers that the denominator is finite for $N \to \oo$,
while the numerator is equivalent to a sum of the type 
$\sum_{n=2}^N n^{(q-3)/2}$, diverging as $N^{(q-1)/2}$ if $q>1$.
This means $\tau = 2$ for RW near a wall, in any $d$. 
The exact result for RW suggests that, in any dimension, for SAW
confined by a wall, the scaling of the contacts
is the same as for RW. Thus, excluded volume effects seem to play no
role in determining $\tau$ for the contacts of the SAW
with a confining wall.

In summary, the contacts discussed above represent a very peculiar example
of scaling random set: in fractal physics we are familiar
with sets in which the number of elements is growing to infinity
together with their average radius of gyration, and criticality implies a
nontrivial scaling for the PDF's of both
quantities. A typical example are percolation clusters~\cite{Stau}, whose size 
PDF has a singular scaling function with nontrivial $\tau$,
like we instead find here for $\Prad$.
In the case considered here the scale invariance of
the set is not accompanied by its number of elements
being broadly, power law distributed. The criticality
is in fact triggered by the length of the whole chain, $N$, becoming
infinite, while $N_c$ remains finite. 
A $\tau$ exponent, possibly super-universal, can also
be defined for the contacts between a SAW and a $d-1$ dimensional
confining wall in the ordinary regime. The case of a RW in presence of wall
can also be treated, yielding the exact classical value $\tau=2$, which 
could apply also to SAW, independent of $d$. 
Exact determinations of these exponents for
SAW in low ($d=2$) or high dimensionality,
and the possible descriptions of the new scalings within the renormalization
group framework remain a challenge for the future.

Partial support from the European Network No. ERBFMRXCT980183 and from
MURST through COFIN-99 is acknowledged. We thank
 U.~Bastolla and E.~Carlon for help and useful discussions.


\vbox{
\begin{table}[!t]
\begin{tabular}{ccccrr}
{\it Lattice} & $\tau$ & $\nu$ & $\nsaw$ & $\Nmin$ & $\Nmax$ 
\\ \hline
2 $d$ & 1.93(2) & 0.76(1) & 3/4~\cite{Nie} & 1000 & 10000
\\ \hline
3 $d$ & 1.51(2) & 0.595(5) & 0.5877(6)~\cite{nu3d} & 1000 & 10000
\\ \hline
FCC & 1.52(2) & 0.594(5) & 0.5877(6) & 3000 & 6000
\\ \hline
4 $d$ & 1.51(3) & 0.52(2) & 1/2 & 1500 & 8000
\\ \hline
5 $d$ & 2.0(2) & 0.49(2) & 1/2 & 1000 & 8000
\\ \hline
6 $d$ & 2.9(1) & 0.49(1) & 1/2 & 1000 & 5000
\end{tabular} 
\caption{Extrapolated values of $\tau$ and $\nu$ for various lattices, using
SAW with length from $\Nmin$ to $\Nmax$.}
\label{T1} 
\end{table}
}

\vbox{
\begin{table}[!t]
\begin{tabular}{cccrr}
{\it Lattice} & $\tau$ & $\nu$ & $\Nmin$ & $\Nmax$ 
\\ \hline
2 $d$ & 1.99(3) & 0.744(5) & 1000 & 6000
\\ \hline
3 $d$ & 1.968(34) & 0.58(1) & 1000 & 8000
\end{tabular}
\caption{As in table \ref{T1}, but for SAW-wall contacts.} 
\label{T2}
\end{table}
}


\end{multicols}

\end{document}